\documentclass[useAMS,usenatbib]{mn2e}
\usepackage{graphicx}
\usepackage{txfonts}

\usepackage{amsmath}

\catcode`\@=11
\def\gsim{\ifmmode{\mathrel{\mathpalette\@versim>}}
    \else{$\mathrel{\mathpalette\@versim>}$}\fi}
\def\lsim{\ifmmode{\mathrel{\mathpalette\@versim<}}
    \else{$\mathrel{\mathpalette\@versim<}$}\fi}
\def\@versim#1#2{\lower 2.9truept \vbox{\baselineskip 0pt \lineskip
    0.5truept \ialign{$\m@th#1\hfil##\hfil$\crcr#2\crcr\sim\crcr}}}
\catcode`\@=12
\def\msun{\hbox{$M_\odot$}}

\def\pn{\par\noindent}

\def\yr-1{\hbox{${\rm yr}^{-1}$}}

\def\log{\hbox{{\rm log}$\,$}}
\def\lb{\hbox{$L_{\rm B}$}}

\def\lr{\hbox{$L_{\rm r}$}}

\def\lsun{\hbox{$L_\odot$}}
\def\lr{\hbox{$L_{\rm r}$}}
\def\zsun{\hbox{$Z_\odot$}}

\def\t9{\hbox{$t_9$}}

\def\zfe{\hbox{$Z_{\rm Fe}$}}

\def\zfesun{\hbox{$Z^\odot_{\rm Fe}$}}
\def\zsun{\hbox{$Z_\odot$}}

\def\pn{\par\noindent}
\def\lbs{\hbox{$L_{\rm B,\odot}$}}

\def\zmax{\hbox{$Z_{\rm max}$}}

\def\m*{\hbox{$M_{\rm stars}$}}

\def\micm{\hbox{$M_{\rm ICM}$}}

\def\zfe{\hbox{$Z^{\rm Fe}$}}
\def\yfe{\hbox{$y_{\rm Fe}$}}
\def\zfes{\hbox{$Z^{\rm Fe}_{*}$}}
\def\zfecm{\hbox{$Z^{\rm Fe}_{\rm ICM}$}}
\def\ho{\hbox{$H_\circ$}}
\def\h50{\hbox{$\ho /50$}}
\def\hst{\hbox{$h_{70}$}}

\begin{document}

\title{
Chemical Evolution on the Scale of
Clusters of Galaxies: \pn A Conundrum? }
\author[Alvio Renzini and Stefano Andreon]{Alvio Renzini$^{1}$\thanks{E-mail: 
alvio.renzini@oapd.inaf.it (AR); stefano.andreon@brera.inaf.it (SA)}
and Stefano Andreon$^{2}$\\ 
 $^{1}$INAF - Osservatorio
Astronomico di Padova, Vicolo dell'Osservatorio 5, I-35122 Padova,
Italy\\
$^{2}$INAF - Osservatorio Astronomico di Brera, via Brera 28, I-20121
Milano, Italy}

\date{Accepted ... 2014; Received April 7, 2014 in original form}
 \pagerange{\pageref{firstpage}--\pageref{lastpage}} \pubyear{2002}

\maketitle
                                                            
\label{firstpage}

\begin{abstract}
The metal content of clusters of galaxies and its relation to their
stellar content is revisited making use of a cluster sample for
  which all four basic parameters are homogeneously measured within consistent
radii, namely core-excised mass-weighted metallicity plus total, stellar
and ICM masses.  For clusters of total mass $M_{500}\simeq
10^{14}\,\msun$ nice agreement is found between their iron content
and what expected from empirical supernova yields. For the same
clusters, there also appears to be at least as much iron in the
 intracluster medium (ICM) as there is still locked into stars (i.e.,
the ICM/stars metal share is about unity). However, 
for more massive clusters the stellar mass fraction appears to drop
substantially without being accompanied by a drop in the ICM
metallicity, thus generating a major tension with the nucleosynthesis
expectation and inflating the metal share to extremely high values
(up to $\sim 6$). Various possible solutions of this {\it conundrum}
are discussed, but are all considered either astrophysically implausible,
or lacking an independent observational support. For this reason we still
entertain the possibility that even some of the best cluster data may
be faulty, though we are not able to identify any obvious
bias. Finally, based on the stellar mass-metallicity relation for
local galaxies we estimate the contribution of galaxies to the ICM enrichment 
as a function of their mass, concluding that even the most massive galaxies must have lost a major fraction of
the metals they have produced.

\end{abstract}


\maketitle

\section{Introduction}
\label{intro}
Clusters of galaxies are the largest bound structures in the Universe,
and have been often regarded as possibly being the best example of a
{\it closed-box} system, i.e., a system
in which all the actors are present from the beginning to the end (e.g., \citealt{white93}).
This is equivalent to assume  that present-day clusters contain,
together with their dark matter, all the baryons in their cosmic share
that have contributed to star formation, all the stars
that have formed out of them and all the metals produced by the
successive stellar generations.  This assumes that the baryonic
fraction of clusters is equal to the cosmic ratio, $\Omega_{\rm
  b}/\Omega_{\rm m}\simeq 0.165$ \citep{komatsu09}, an hypothesis that can be subject to
observational test, and appeared to be verified at least for the richest
clusters \citep{gonzalez07,pratt09,andreon10,leauthaud12}. This assumption clearly fails at
least for groups and low-mass clusters with mass less than $\sim
10^{13}\,\msun$, whose gas content
can be much lower than the cosmic share, indicating that
baryons may have been lost by these system or never incorporated
in them (e.g., \citealt{renzini93,mcgaugh10}).

To the extent that the closed box assumption is close to reality, clusters of galaxies can offer a unique opportunity to study
chemical evolution on the largest scale for which the census of all
the mentioned components is virtually complete\footnote{\footnotesize
  The largest possible scale for chemical evolution studies in the
  Universe as a whole, but the current census of baryons and metals in
  the general field  is  incomplete at all redshifts.}, hence allowing us to obtain an empirical
measure of the chemical yield(s) ($y$) and  the fraction of
cosmic baryons turned into stars, i.e., the {\it efficiency} of galaxy
formation. 

In this paper we revisit these issues using updated cluster data that
we consider of the best quality for our purposes, i.e., for which
  total, stellar and ICM mass and metallicity have been 
    homogeneously measured
  within consistent radii. The paper is organized
as follows: in Section \ref{history} we summarize the basic understanding of cluster chemistry prior of the newer data presented
in Section \ref{data}, which are then elaborated in Section
\ref{results}. Section \ref{implications} expands on the implications
of the new results which appear to pose new
challenges, an apparent {\it conundrum} where in the most massive
clusters there appears to be much more iron than the cluster galaxies
may have reasonably produced. Possible solutions to this conundrum are then listed
and discussed in Section \ref{conundrum} whereas Section \ref{winds}
presents a semi-empirical estimate of the amont of ejected metals as a
function of present-day galaxy mass. Finally, our conclusions are
summarized in Section \ref{end}

\section{Basic Cluster Chemistry}
\label{history}
It is known since a long time that the
abundance of iron in the  intracluster medium (ICM) is nearly constant at the level of $\sim
0.3Z^{\rm Fe}_\odot$ (e.g., \citealt{arnaud92}), at least for clusters
whose ICM is hotter than $\sim 2$ keV.
Based on literature data, it was then inferred that  
similarly  constant is  the iron-mass-to-light
ratio (IMLR),
defined as the total mass of iron in the ICM over the total stellar luminosity
of the whole cluster \citep[][hereafter GR11]{renzini97,renzini04,gr11}. 

Before presenting and discussing new cluster data which may change
this picture, for sake of
comparison we synthetize here the major conclusions reached in the
above references. 
Thus, following GR11, we have for the IMLR
of the ICM:

\begin{equation}
({\rm Fe}M/\lb)_{\rm ICM} = \zfecm{\micm\over\lb} \simeq 0.010\,h_{70}^{-1/2} ,  
\label{imlricm}
\end{equation}
where it was adopted $\zfecm=0.3\zfesun$, and:
\begin{equation}
\micm/\lb\simeq 25h_{70}^{-1/2}\quad (\msun/\lsun),
\label{micmlb}
\end{equation}
for the mass of the ICM, a value that had been  derived for the Coma cluster \citep{white93}.
We had also taken $\zfesun=0.00124$ for the photospheric iron abundance
\citep{asplund09}, virtually identical to  $\zfesun=0.00126$ given
by \cite{anders89}  for the {\it meteoritic} iron
  abundance. However, X-ray studies typically refer to the solar
 {\it photospheric} iron abundance, for which Anders \& Grevesse give
 $\zfesun=0.0018$, the value we adopt here as unit for the ICM
 abundances while we keep the Asplund et al.
value as unit for the stellar abundances.
This means that in the ICM there are $\sim 0.015\,\msun$  of iron for each
solar luminosity of the cluster galaxies. Iron is also locked into
stars and galaxies, and assuming that  the average iron abundance of stars
is solar the IMLR  of cluster galaxies is then:

\begin{equation}
({\rm Fe}M/L)_{\rm gal}=\zfes{\m*\over\lb}\simeq 0.006\quad 
(\msun/\lsun),
\label{fegal}
\end {equation}
adopting $\m*/\lb=5.3\, (\msun/\lbs)$ from the population models of
\cite{maraston05} for a \cite{kroupa01}
IMF, having assumed an age of 11 Gyr and average solar
metallicity,  as appropriate for the early-type galaxies that contribute the bulk of stellar mass in clusters.
Thus, the total, cluster IMLR, sum of ICM and galaxies IMLRs, is:
\begin{equation}
\begin{tabular}{ r l }
\(  ({\rm Fe}M/L)_{\rm  cl} $ &  \( \simeq 6\times 10^{-4}\times (\micm/\lb)
     h_{70}^{-1/2} \)
\\
 & \(  +1.2\times 10^{-3}{\m*\over\lb} \simeq 0.021 \quad  (\msun/\lsun), \)
$$
 \end{tabular}
\label{feclust}
\end{equation}
for $\hst=1$. Notice again that here and the following we distinguish
between the solar iron used for the stars, which is the
photospheric iron from from Asplund et al. (2009), and the solar
iron used for the ICM, which is the photospheric iron from
Anders \& Grevesse (1989).
Quite an interesting quantity is the {\it iron share} between ICM and
galaxies, i.e., the ratio of the iron mass in the ICM over that  in galaxies:
\begin{equation}
{\zfecm\micm\over\zfes \m*} \simeq 2.5\times h_{70}^{-1/2},
\label{share}
\end{equation}
This means that there is al least as much  mass of iron
diffused in the ICM as there is still locked into stars, indicating that galaxies 
lost at least as much  iron as were able to retain into their stellar
populations. We refrain from attaching precise uncertainties to these estimates, as they depend
on the few explicit assumptions that have been made above, such as the  $M/L$ ratios.

Next issue is whether our current unerstanding of stellar
nucleosynthesis is able to account for the huge mass of iron contained
inside clusters of galaxies. Following again GR11, we introduce the
supernova productivity factors, respectively $k_{\rm CC}$ and $k_{\rm
  Ia}$ for core collapse (CC) and Type Ia supernovae, which give the
number of SN event produced per unit mass of gas turned into
stars. For a ``Salpeter-diet'' IMF\footnote{The slope $s$ of a
  Salpeter-diet IMF is 2.35 above $0.5\;\msun$ and flattens to 1.35
  below. It is virtually identical to the IMF proposed by
  \cite{chabrier03} or \cite{kroupa01}} $k_{\rm CC}$ ranges from $\sim 5\times 10^{-3}$
to $\sim 10\times 10^{-3}$ (events for every $\msun$ of gas turned
into stars), depending on the assumed minimum stellar mass for
producing a CC event. The above values refer to 12 $\msun$
and 8 $\msun$ for such minimum mass, respectively, and in the following we
adopt $k_{\rm CC}=7\times 10^{-3}$.

The SNIa productivity is more difficult to estimate, with values
ranging from $k_{\rm Ia}\simeq 10^{-3}$ (events/$\msun$) to $2.5\times
10^{-3}$, depending on the semi-empirical method used to derive
it form observed SNIa rates (GR11).  Given this large uncertainty,
in the following we consider this full range of $k_{\rm Ia}$.

The bulk of iron produced by supernovae comes from the decay of the
radioactive $^{56}$Ni which mass per event can be estimated from the
SN light curve.  For both kinds of SNe the $^{56}$Ni mass varies
greatly from one event to another. Averaging over many events one has
$<\! M(^{56}{\rm Ni})\!>_{\rm CC}=0.057\,\msun$ (\citealt{zampieri07})
and $<\! M(^{56}{\rm
  Ni})\!>_{\rm Ia}=0.58\,\msun$ (\citealt{howell09}), respectively for CC and Type Ia
supernovae (cf. GR11). We adopt here $<\!
M({\rm Fe})\!>_{\rm CC}=0.07\,\msun$ and $<\! M({\rm Fe})\!>_{\rm
  Ia}=0.7\,\msun$, having allowed for a modest contribution from
direct production of iron in the SN explosion, ejected as such rather
than as $^{56}$Ni. So, every 1,000 $\msun$ of gas turned into stars, CC
and Type Ia supernove produce $7\times 0.07\simeq 0.5\,\msun$ and
$(1-2.5)\times 0.7\simeq (0.7-1.7)\,\msun$ of iron, respectively. Together, they
produce $(1.2-2.2)\,\msun$ of iron, with the major uncertainty coming
from the semi-empirically estimated productivity of Type Ia supernovae
($k_{\rm Ia}$). Thus, in solar units the iron yield is
expected to be  in the range:
\begin{equation}
\yfe\simeq (1-2)\zfesun.
\label{yield}
\end{equation}

This iron yield needs to be converted into a IMLR in order to be
compared to the value measured in rich clusters of
galaxies. To this end one needs to estimate what is the present $B$-band
luminosity of a stellar population resulting from the conversion into
stars of 1,000 $\msun$ of gas. We
assume again the bulk of stars in clusters to be 11 Gyr old, hence  $\m*(11)/\lb(11)=5.3$.  However, this refers to the current mass of
the population, which compared to the initial mass has been  reduced by the
mass return. The same \cite{maraston05} models for a \cite{kroupa01}
IMF give
$\m*(11)=0.58\times \m*(0)$. So, the initial-mass to present-light ratio is
$\m*(0)/\lb(11)= 5.3/0.58=9.14\; \msun/\lbs$. Hence, the $B$-band
luminosity of a stellar population of initially 1,000 $\msun$ is
$1,000/9.14 = 109\;\lbs$. Finally, the {\it predicted} IMLR is
therefore $\sim (1.2-2.2)/109= (0.011-0.020)\; \msun/\lbs$, 
  falling just marginally short of the measured value in clusters, i.e., $\sim
0.021\,\msun/\lbs$.  However,  the value of $<\! M(^{56}{\rm
  Ni})\!>_{\rm CC}$ adopted above actually pertains only to the
SNII-Plateau type  of CC supernovae, which is equivalent to ignore the
contribution of stars more massive than roughly $40\,\msun$. Possible
contributions, if any, 
from other CC types (SNIb, SNIc), pair instability supernovae
\citep{heger02} and hypernovae \citep{nomoto13} can only ease this
marginal mismatch.
Assuming the younger age of 9 Gyr for the bulk of
stars in clusters, then one would have $\m*(9)/\lb(9)=4.2$ and repeating
the same calculation [from Equation \eqref{feclust}] on one would get a predicted IMLR=(0.009--0.016),
still  marginally consistent with the observed value, within the combined
uncertainties. Therefore, the result is not strongly dependent on
the assumed age of stars in clusters.

This was the reassuring conclusion in GR11: a standard IMF
(Salperter-diet, Kroupa or Chabrier), coupled to our best current
understanding of iron production by CC and Type Ia supernovae, account
reasonably well for the observed amounts of iron in clusters of
galaxies, which is partly diffused in the ICM, partly locked into
stars. 

Such an optimistic view was further reinforced considering, besides
iron, also oxygen and silicon which are predominantly produced by CC
supernovae and therefore their yield is much less sensitive to the
uncertainty affecting the productivity of Type Ia supernovae ($k_{\rm
  Ia}$).
Using standard nucleosynthesis prescriptions GR11 (see also
\citealt{renzini04}) derive the predicted oxygen-mass-to-light ratio
and the silicon mass-to-light ratio for a $\sim 11$ Gyr old
population,
as a function of the IMF slope between 1 and 40 $\msun$. Such
predicted ratios are then almost independent of the IMF for
$M<1\;\msun$, and  can be approximated as:
\begin{equation}
{\rm log}\, M_{\rm O}/\lb \simeq -1.13 -1.37(s-2.35)\quad
(\msun/\lbs),
\end{equation}
and
\begin{equation}
{\rm log}\, M_{\rm Si}/\lb \simeq -2.07 -1.27(s-2.35)\quad
(\msun/\lbs),
\end{equation}
which, for the Salpeter slope $s = 2.35$,  give values in excellent agreement with those observed in clusters of galaxies: respectively
$\sim 0.1$ and $\sim 0.01\,\msun/\lbs$ for oxygen and silicon (e.g.,
\citealt{finoguenov03}).

It is worth emphasizing that  these conclusions rest on some relatively old cluster data from
the literature, and as a cautionary point GR11 mentioned that ICM
mass, cluster light and abundances were often drawn from different
sources which occasionally  might have used different cluster sampling
for different quantities (e.g., within $r_{200}$ or $r_{500}$ or
whatever).
To hopefully overwhelm these limitations, in the next sections we
re-asses these issues by making use of  recent cluster data for
  which all these quantities have been obtained within consistent radii.
Such possibly more homogeneous and reliable  cluster measurements may suggest that
Nature is more complicated
than in the reassuring picture summarized above. On the other hand,
such optimistic scenario has been recently questioned by
\cite{loewe13} according to whom the cluster metals would exceed by a
factor of $\sim 2-3$ the expectations from nucleosymthesis. The
discrepancy does not arise from different adopted supernova yields, as
this author adopted basically the same prescriptions as done here from GR11. It arises instead from 
different cluster parameters, specifically from the stellar mass
fraction $\m*/(\m* +\micm)$ which was taken here to be $\sim 0.17$
as opposed to $\sim 0.1$ in \cite{loewe13}, where however a value $\ge
0.25$ is not excluded.

\begin{table*}
\caption{$\quad\quad\quad$ Cluster Properties}\label{tab:table1}
\begin{tabular}{llllllllll}
\hline
\hline
Clusters ID&  $z$ &  log$(M_{500})$& Error& log$(L_{\rm r})$&  Error &
$f_{\rm gas}$&   Error & $Z^{\rm Fe}/Z^{\rm Fe}_\odot$ & Error\\
\hline

A1795      &0.062 & 14.78 & 0.04 & 12.02 & 0.05 & 0.104 & 0.006  & 0.22 & 0.06\\
A1991      &0.059 & 14.09 & 0.06 & 11.82 & 0.12 & 0.094 & 0.010  & 0.40 & 0.09\\
A2029      &0.078 & 14.90 & 0.04 & 12.36 & 0.08 & 0.123 & 0.007  & 0.30  & 0.10\\
MKW4       &0.020 & 13.89 & 0.05 & 11.61 & 0.14 & 0.086 & 0.009  & 0.35 & 0.05\\
3C442A    &0.026 & 13.59 & 0.03 & 11.33 & 0.14 & 0.068 & 0.006  & 0.26 & 0.05\\
NGC4104  &0.028 &13.69 & 0.05 & 11.44 & 0.20 & 0.069 & 0.009  & 0.30 & 0.07\\
A160         &0.045 &13.90 & 0.06 & 11.69 & 0.12 & 0.085 & 0.009  & 0.33 & 0.11\\
NGC5098  &0.037 & 13.30 & 0.07 & 11.47 & 0.15 & 0.108 & 0.021  & 0.23 & 0.04\\
A1177       &0.032 & 13.72 & 0.06 & 11.41 & 0.15 & 0.060 & 0.009  & 0.22 & 0.05\\
RXJ1022+383 &  0.054 & 13.90 & 0.07 & 11.79 & 0.16 & 0.075 & 0.007  & 0.28 & 0.09\\
A2092       &0.067 & 13.95 & 0.08 & 11.70 & 0.10 & 0.078 & 0.013  & 0.39 & 0.20\\
NGC6269  &0.035 & 13.93 & 0.09 & 11.75 & 0.12 & 0.076 & 0.011  & 0.27 & 0.07\\
\hline
\end{tabular}
\end{table*}

\section{Cluster Data}
\label{data}

The global basic parameters of clusters of galaxies (namely total
  mass, ICM mass, stellar luminosity or mass, and metallicity) have
  been measured for many clusters over the last decades. Yet, most
  often at least one of these four parameters is missing and
  furthermore these quantities may have been measured within different 
radii in different studies. Culling cluster samples from different
sources is then prone to significantly increase the scatter in any
relation among these four quantities. For example, mass estimates
derived by different authors may systematically differ by up to $\sim
45\%$  \citep{rozo14}. The lack of uniform X-ray analysis of the various cluster
samples is indeed one of the limiting factors for studies of the
cluster scaling relations \citep{bender14}.

In spite of these limitations, three main trend with the cluster
  total mass are well documented in the literature, namely: \par
1) the gas
  fraction of clusters ($M_{\rm gas}/M_{\rm total}$) moderately
  increases with cluster mass (e.g., using total mass measurements: \citealt{vikhlinin06,
    arnaud07, gastaldello07, ettori09, sun09, andreon10, gonzalez13};
  and using ICM temperature or other  proxies to mass, e.g.,
  \citealt{grego01, sanderson03, giodini09}). \par
2)
  The cluster metallicity is constant with cluster mass (e.g., \citealt{
    sun12, vikhlinin05}, or using temperature as a proxy
  to mass, e.g., \citealt{matsushita11, balestra07,  andreon12b}).

\par
3) the
  stellar mass fraction ($M_{\rm star}/M_{\rm total}$) decreases with
  cluster mass (e.g., \citealt{andreon10,andreon12a, gonzalez13, leauthaud12,
kravtsov14, lin12}).

To exemplify and best quantify these trends, after
  extensive explorations of the existing literature we have identified
  just 12 clusters for which all four quantities have been measured
  within consistent radii, with total and ICM masses having been
  derived from X-ray data assuming hydrostatic equilibrium.  To our
  best knowledge there are no other samples of clusters with these
  four quantities have been homogeneously measured within consistent
  radii.

This {\it primary} sample is formed by the
subsample of relaxed clusters among those with accurately
measured masses within the $r_{500}$ radius\footnote{$r_{\Delta}$ is the
radius within which the enclosed average
mass density is $\Delta$ times the critical density.} ($M_{500}$) in \cite{vikhlinin06}
and \cite{sun09}. These masses were derived from X-ray surface
brightness and temperature profiles, assuming spherical symmetry, hydrostatic equilibrium
and purely thermal pressure (i.e., ignoring turbulence and magnetic
field contributions to pressure). Such clusters were also selected for
lying within the area covered by the  SDSS and with $z<0.05$.
Clusters in this primary sample are listed in Table 1.

Besides $r_{500}$ and $M_{500}$, \cite{vikhlinin06} and \cite{sun09}
have also measured 
concentrations $c_{500}$ and gas masses $M_{gas,500}$, as a result of
the same best fit of the X-ray surface brightness and temperature profiles. 
Cluster masses and gas fractions ($M_{\rm gas,500}/M_{500}$) reported in Table 1 are deprojected
values within the sphere of radius $r_{500}$. By construction, this mass derivation makes no
prior assumption on the mass profiles, which instead are derived
directly from the X-ray data. The parent sample of \cite{vikhlinin06}
and \cite{sun09}, although 
not complete in mass, is considered  to be representative of the
general cluster population, having  been successfully used to calibrate the mass-$T_{\rm X}$
scaling relation for cosmological estimates (e.g. \citealt{vikhlinin09}).

\begin{figure}
\includegraphics[width=84mm]{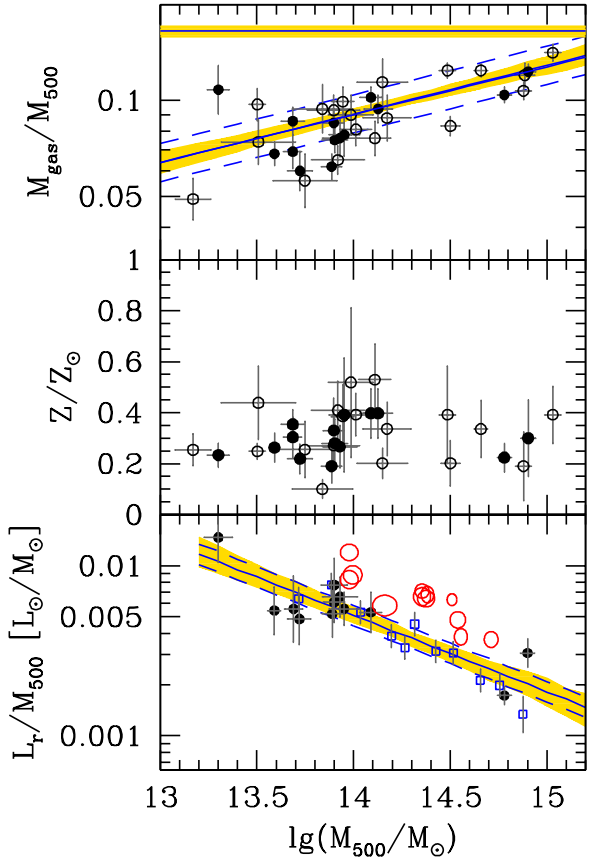}
\caption{Various quantities are plotted as a function of the total
cluster mass within $r_{500}$. Filled circles refer to our sample of
12 clusters from Table 1, with $M_{\rm gas}$ and
$M_{500}$ from Vikhlinin et al. (2006) and Sun et al. (2009), 
metallicity $Z$ from Vikhlinin et al. (2005) and Sun et al. (2012) and $L_{\rm r}$
from Andreon (2012a). All these four parameters are measured
homogeneously within consistent radii. Open circles refer to
clusters from the same sources, but  
for which no $L_{\rm r}$ measurements are available. {\it Top panel}: the cluster gas
fraction with the corresponding best-fit linear relation. The
horizontal line is drawn at the level of cosmic
baryon fraction = 0.165. {\it Middle Panel}: The ICM iron abundance. {\it Bottom Panel}: The cluster $L_{\rm r}/M_{500}$
ratio. Open squares refer to clusters
from Andreon (2012a) for which ICM mass and metaliicity have not been
measured in Vikhlinin et al. (2006) and Sun et al. (2009). These are typically stacks of 5 clusters. The small red ellipses refer to the clusters
from Gonzalez et al. (2013) with their axes corresponding to  1$\sigma$
error bars. For the offset
affecting the Gonzalez et al. sample see the main text.
 In the top and bottom panels the best-fit
relations to the 12 clusters and their uncertainty range are also shown, with the yellow
band representing the 1-$\sigma$ error from the best fit and the blue
dashed lines the best fit linear relation $\pm$ the 1-$\sigma$ intrinsic scatter }
\label{goods}
\end{figure}

Based on the very same X-ray data,  cluster metallicities  have
been derived by \cite{sun12} and \cite{vikhlinin05} via
spectral fitting and their results are also reported in Table 1. 
These metallicities are {\it mass-weighted} over the whole ICM, and are
typically $\sim 30\%$ lower than the luminosity-weighted ones that
come directly from the X-ray spectral fits. Hence, unlike
luminosity-weighted ones they are not affected by the central
abundance enhancement, which is typically $\sim 5\%$ in mass within
  $r_{500}$, or $\sim 10\%$ within $r_{2500}$
(e.g. \citealt{degrandi04}). For this reason, we adopt deprojected 
mass--weighted metal abundance between $0.3<r/r_{500}<0.6$ for clusters
in \cite{sun09} and in A2092 
(the latter from M. Sun, private
communication).  Two clusters (A1795 and A2029) in our
primary sample lack a
mass-weighted metallicity measurement from  \cite{sun09} and \cite{vikhlinin05}.  For A1795 we adopt from 
\cite{vikhlinin06} the
mean abundance measured outside the central Fe enhancement region, at
$r\approx 0.45 r_{500}$ . 
For  A2029 we adopt 0.3 solar for the iron abuncance  outside its central cool
core (\citealt{aaron02} and
references therein).
The average iron abundance of the ICM of the 12 clusters in Table 1 is therefore
$\zfecm\simeq 0.3\zfesun$. 
These abundances are in the \cite{anders89} scale  where $\zfesun=0.0018$, 
and have been updated for the recent change of the Chandra calibration, an
upward (abundance) change by a factor 1.12 \citep{andreon12b}.
Iron abundance data exist for a much larger sample of clusters
  than reported on this figure, showing that at least above $\sim 2$
  keV the iron abundance is independent of ICM temperature, hence
  mass (e.g., \citealt{andreon12b}, and references therein). Thus,
  an abundance $\sim 0.3$ solar appears to be applicable to all
  massive clusters studied so far.

The $r$-band luminosity, $\lr$ within the radiii $r_{500}$ and $r_{200}$
has been derived in \cite{andreon12a} using SDSS data
by integrating the luminosity function of the
red cluster galaxies, adding up the luminosity of the BCG and the galaxy
light below the detection threshold (by extrapolating the luminosity
function, see \citealt{andreon10} and \citealt{andreon12a} for
details). The $\lr$ value of A2029 reported in Table 1 has been
recomputed from CFHT MegaCam images, strictly following the same procedure, because SDSS
data turned out to be partially corrupted at the sky
location of A2029 (S. Andreon, in preparation).
Red galaxies are those
within 0.1 redward and 0.2 blueward in $g-r$ with respect to the red
sequence in  the colour--magnitude plot of each individual cluster. The
intracluster light is negligible within such large radii
(\citealt{zibetti05}; \citealt{andreon10}; \citealt{giallongo13}; \citealt{presotto14})
once one accounts for  the light emitted by undetected galaxies, by
the outer regions of galaxies and by the BCG, which are all included
in the present estimates of $\lr$. As for all other quantities, $\lr$ is the deprojected
value within the sphere of radius $r_{500}$ (or $r_{200}$). For this
deprojection the cluster light is assumed to follow a NFW distribution
\citep{NFW}.

We supplement this sample of relaxed clusters, having both top-quality
X-ray data and $L_{\rm r}$ measurements,  with other cluster samples with less
complete datasets but which help establishing the main trends. These
{\it secondary} samples include either relaxed cluster missing measurements of the
optical luminosity, or clusters with masses derived from the caustic
technique without any restiction on their dynamical status. 
Also for these samples, measurements are performed within
consistent radii.
Thus, 
we also consider the clusters in \cite{vikhlinin06} and \cite{sun09} 
lacking a measured $\lr$ but which  are useful
to delineate the mass dependency of the gas fraction and metallicity.
Besides them, we also consider clusters  from \cite{andreon10} with high quality $\lr$, but lacking
high-quality X-ray data. This
supplementary sample consists of 54 clusters with dynamical 
(caustic) masses derived from their estimated escape velocity  with a
typical $0.15$ dex error (\citealt{rines06}).
These clusters are selected independently of their dynamical status,
and their masses, unlike those derived from X-ray data, 
do not assume hydrostatic equilibrium and the relaxed status of the cluster. On the other
hand, these measurements make a (weak) assumption about the velocity
dispersion anysotropy. The $r$-band luminosity of these
clusters have been derived as for the
primary sample, except
that measurements were performed only within  $r_{200}$
\citep{andreon10}. This sample  helps to delineate the
  mass-dependency of $\lr/M_{500}$, hence of the stellar fraction,
  confirming the trend derived from the 12 primary clusters.

Finally, we use also the cluster sample from \cite{gonzalez13},  where
core-excised X-ray
temperatures were measured  and total  masses were derived  assuming a mass vs X-ray temperature
relation. 
From their
deprojected stellar masses we removed the $M/L_{\rm I}$ ratio  assumed by the authors
(2.65), and we then converted the resulting $I$-band luminosities into $r$-band
ones using \cite{maraston05}  models for a Kroupa IMF, solar
metallicity, 11 Gyr old simple stellar population. 
No iron abundances are available for these clusters.
\section{Results}
\label{results}

The top panel of Figure~\ref{goods} shows the gas fraction as a
function of the cluster mass for the clusters in 
\cite{vikhlinin06} and \cite{sun09}; filled circles refers to objects
in our primary sample. The corresponding best
fit relation  is: 
\begin{equation}
\log \frac{M_{gas,500}}{M_{500}} \simeq  (0.15\pm0.03)  (\log
M_{500}-14.5) -0.97\pm 0.02. 
\label{eq2}
\end{equation}
There is a modest increase of the gas fraction with cluster mass, as
already pointed out in the literature
\citep{sun09,andreon10,gonzalez13}.  
After accounting for observational errors the intrinsic
scatter is also small ($\sim 15$ \%), indicating minor 
cluster to cluster stochasticity in gas fractions (\citealt{andreon10}).

The middle panel of Figure \ref{goods} shows the 
observed iron abundances in the ICM (mass weighted).  There is no
appreciable trend of iron abundance with cluster mass, i.e.,
$\zfe\simeq$ constant, apart from a hint for a possible local maximum
around  $M_{500}=10^{14}\,\msun$,
resulting from an apparent maximum in clusters with $kT\simeq
2$ keV which may be spurious \citep{renzini97}. This virtually constant
iron abundance is found also among the 130 clusters  (mostly with $kT>3$ keV) in
\cite{andreon12b}. For these clusters the iron abundances are still
luminosity-weighted  and non-core excised.

Finally, the bottom panel of Figure~\ref{goods} shows the cluster light-to-mass ratio $\lr/M_{500}$:
the cluster $r$-band luminosity is not proportional to the
cluster mass, but its growth is much slower, i.e., massive clusters emit less
luminosity per unit cluster mass than less massive clusters 
\citep{andreon10,andreon12a,gonzalez13}. As in the rest of the figure, the solid circles in the 
the bottom panel refer to  our primary sample, with data being best fitted
by: 
\begin{eqnarray}
\log \frac{L_{\rm r,500}}{M_{500}} = -(0.45\pm0.08)({\rm
  log}M_{500}-14.5)-2.51\pm 0.05,
\label{eq3}
\end{eqnarray}
which is also reported in the bottom  panel. This best fit relation
provides also an 
excellent match to other datasets, namely the full sample from
\cite{andreon10}  (the open squares in Figure \ref{goods}, where all quantities are however measured within
$r_{200}$  instead of within  $r_{500}$) and the sample from
\cite{gonzalez13} (red ellipses in Figure \ref{goods}). 
Notice that for the former set of clusters the open squares in Figure
\ref{goods} refer to the stack in groups up to five each, so  to
reduce the scatter given the larger errors.
The offset of the latter set of clusters seen on the bottom panel of Figure
\ref{goods} compared to the other clusters may be partly due to a differential selection effect.
\cite{gonzalez13} clusters are indeed drawn from an optically-selected sample and have a very
dominant galaxy (BCG) contributing up to 40\%  to the total luminosity, 
thus are not selected independenty of the quantity being measured (the
cluster optical luminosity). Instead, the other clusters in Figure
\ref{goods}  are selected in X-ray, hence  independenty of their optical
luminosity.  Moreover, part of the offset may well be due to
the known systematic differences in mass estimates \citep{rozo14,applegate14,linden14}, which once
more illustrates the need for using homeogeneously
derived quantities  as much as possible. 

All in all, once a vertical/horizontal offset is allowed to account for
selection effects the three sets of clusters (75 clusters in total)  appear to define a tight
$\lr -M$ relation, with the $\lr/M$ ratio dropping by a factor $\sim
3$ for cluster masses between $10^{14}$ and $10^{15}\,\msun$.

We recall that for clusters shown as circles in Figure
\ref{goods} $M_{500}$ was measured from X-ray data assuming hydrostatic
equilibrium, whereas for clusters shown as open squares $M_{200}$ was
measured with the caustic method and those shown as ellipses came from
the $T_{\rm X}-M$ relation.  The agreement  of the $\lr/M_{500}$ ratios (proportional to the stellar mass fraction) vs mass
relations among these various datasets  (apart from the mentioned
offset of Gonzalez et al. clusters) suggests that these mass
measurements  are consistent with each other.  Although with larger
errors compared to hydrostatic modelling (typically $\sim 0.15$ dex vs
$\sim 0.05$ dex) this in particular applies to masses derived with the
caustic method. 

At variance with these results, \cite{russi} find an almost linear
$\lr -M_{500}$ relation, hence a stellar mass fraction almost constant
with total cluster mass. When measuring $\lr$ in the same way as in
\cite{andreon10} (i.e., within $r_{200}$) \cite{russi} find the same
results as \cite{andreon10},
so they ascribe the discrepancy to the use of caustic masses for
$M_{200}$.  However, in \cite{russi} cluster  masses were not measured directly, but estimated from the cluster richness.
One possible origin of the discrepancy is the large scatter of the
richness--mass relation and the  contamination affecting especially  the lower mass
clusters, leading to overestimate their total mass, hence to
underestimate their stellar fraction (\citealt{kravtsov14}).
In any event, a decreasing trend of the stellar mass to halo mass
ratio, similar to that shown in Figure \ref{goods}, is found in many
studies dealing with the efficiency of baryon-to-stars conversion as a
function of halo mass (e.g., \citealt{leauthaud12,behroozi13,kravtsov14,birrer14}).

\begin{figure}
\includegraphics[width=70mm]{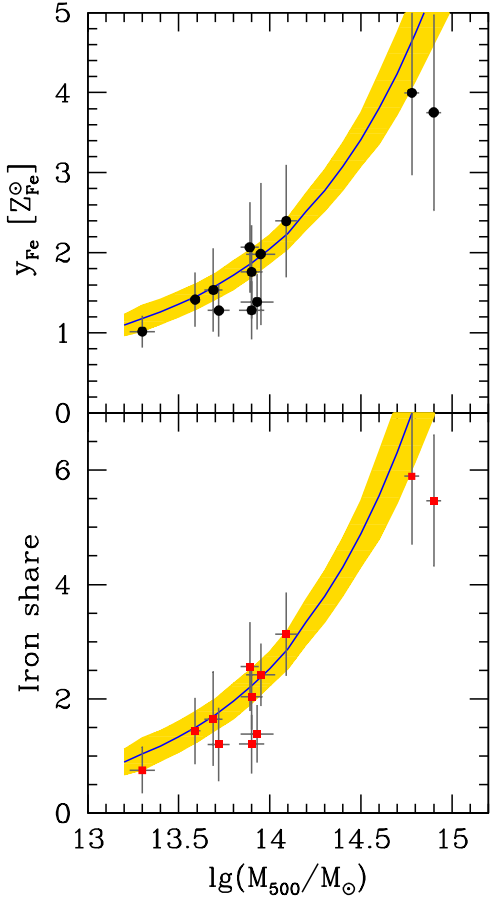}
\caption{{\it Upper Panel:} the apparent iron yield for the 12 clusters in
  Table 1 as a function of $M_{500}$, using
Equation \eqref{yield2}. The relation  implied together by Equation \eqref{eq2} and
Equation \eqref{eq3} is also shown, with the yellow error band
corresponding to those shown in Figure \ref{goods}, and assuming an
iron abundance 0.3 solar (the average of the data shown in Figure
\ref{goods}). {\it Lower Panel:} the corresponding iron share, i.e.,
the ratio of iron mass in the cluster ICM over that in galaxies.
Notice that these yields are in units where $\zfesun = 0.00124$.}
\label{yields}
\end{figure}

\section{Implications}
\label{implications}
The striking implication of combining the information in the three panels of
Figure \ref{goods} is that the total mass of iron in the ICM =
$\zfecm*M_{gas,500}$ increases with cluster mass  while the
$\lr/M_{500}$ ratio (i.e., the stellar mass fraction)
decreases. Taking Figure \ref{goods} at face value, such a decrease is
a factor of $\sim 10$ over 2 dex in $M_{500}$, and yet the iron abundance in the ICM remains
the same ($\sim 0.3\times$ solar) in spite of the drop in the 
mass fraction of the stars which should have produced such mass of iron. 
In more precise quantitative terms, one can estimate the empirical
iron yield as:
\begin {equation}
\yfe=\zfesun{{\m* + 1.45(\zfecm/\zfesun) \times M_{\rm gas}}\over \m*(0)},
\end{equation}
where it is assumed that the average iron abundance in the cluster
stars is solar and $\m*(0)$ is the
mass of gas that went into stars whose present mass is now reduced to
$\m*$ by the mass return from stellar mass loss. This equation can be
further elaborated into:
\begin {equation}
\yfe=\zfesun {\m*\over \m*(0)}\bigl(1+1.45
{\zfecm\over\zfesun}\times{M_{\rm gas}\over\lr}\times {\lr\over
  \m*}\bigr),
\label{yield2}
\end{equation}
where $\zfesun$ in front of these expression is the photospheric iron
from \cite{asplund09},  $\zfecm/\zfesun$ comes from Table 1 and
the factor 1.45 is the ratio of the photospheric solar iron abundance
from \cite{anders89}  over that from \cite{asplund09}.
We then adopt for the residual to initial mass ratio $\m*/\m*(0)=0.58$
and $\m*/\lr=3.24$, as appropriate for a solar metallicity, 11 Gyr old
simple stellar population with ``Krupa IMF'', as  from the synthetic models of
\cite{maraston05}. These expressions for the yield
assume that all clusters have evolved as closed systems as far as
metals and stars are concerned, which may not be the case (see below). 

\begin{figure}
\includegraphics[width=84mm]{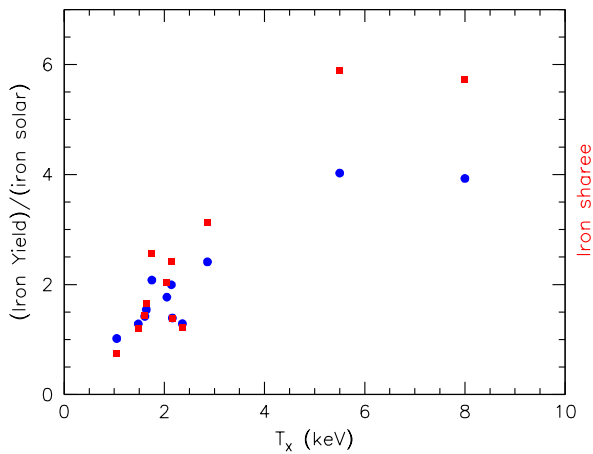}
\caption{The iron yield (blue circles) and share (red squares) as in Figure \ref{yields} but as a function of cluster temperature for the clusters in Table 1 }
\label{yieldt}
\end{figure}

Figure \ref{yields}  shows the
resulting empircal yields having fed data in Table 1 into Equation
\eqref{yield2}. This {\it apparent} yield is in the range between $\sim
1$ and 2$\times\zfesun$ for clusters with total mass up to $\sim 10^{14}\,\msun$, just as expected from the semiempirical
estimates summarized in Equation \eqref{yield}. However, for the two
most massive clusters of the sample the iron yield turns out to be
$\sim 3$ times solar, with a rising trend with cluster mass which is
primarily driven by the decreasing trend in $\lr/M_{500}$ (hence
stellar mass fraction) shown in
Figure \ref{goods}.
Much of these trends is due to very low $\lr$ luminosity
for their total mass of the two most massive clusters and one may
argue that these two clusters may be exceptional, or that some of their
parameters were erroneously estimated.
The fact is, however, that these two clusters follow the general trend
shown in Figure \ref{goods}, hence they do not
appear to be exceptional outliers. Indeed, the \cite{gonzalez13} and
\cite{andreon10} clusters nicely fill the gap between the lower mass
clusters and the two most massive ones in our primary sample. 
The same data are shown in Figure \ref{yieldt} as a function of ICM
temperature, where the (possibly spurious) bump at $T_{\rm X}\sim 2$ keV is also apparent.

We emphasize that the increasing iron yield with cluster mass is the
combined result of a decreasing $\lr/M_{500}$ and a constant
metallicity. For deriving the $\lr/M_{500}$ trend  we use  three independent
measurements \citep{gonzalez13,andreon10,andreon12a} whereas the
constancy of metallicity is well documented in the literature for large cluster samples
(e.g. the 130 clusters in \citealt{andreon12b}).  Thus, this trend is driven by a large sample of
data, not just by the two most massive clusters clusters in our
primary sample. 

Figures \ref{yields}  and \ref{yieldt} show also the {\it iron share} between the ICM
and galaxies, as from Equation \eqref{share}, once more assuming that the
average metallicity of the stars is solar. Again, for clusters up to
$\sim 10^{14}\,\msun$ the iron share is between $\sim 1$ and 2, but for
the two most massive clusters it appears to be much higher,
around  $\sim 4$, i.e., there
appears to be $\sim 4$  times more iron out of galaxies than within them!

\begin{figure}
\includegraphics[width=84mm]{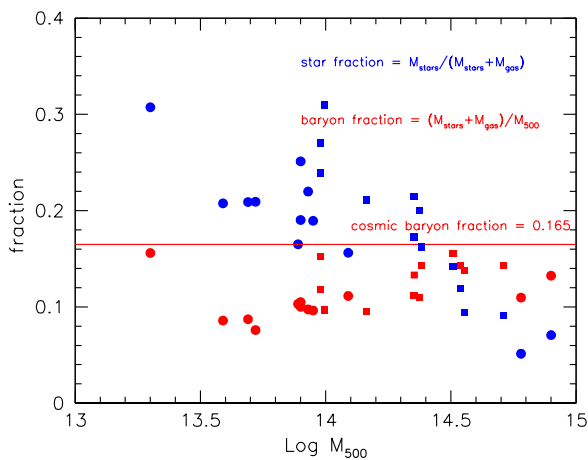}
\caption{The star fraction (blue) and the baryon fraction (red) of the
  clusters in our primary cluster sample (filled scircles) and in the
  sample of Gozalez et al. (2013) (filled squares)}
\label{fractions}
\end{figure}

Figure \ref{fractions} shows in red the baryon fraction of the clusters,
i.e., $(\m* +M_{\rm gas})/M_{500}$, both for our primary cluster
sample (filled circles) and for the sample of \cite{gonzalez13}
(filled squares).  Notice that this bayon fraction lies systematically below the
cosmic fraction 0.165 and appears to increase slightly with
  cluster mass (ignoring the least massive cluster)  as
generally found (e.g., \citealt{andreon10,leauthaud12, gonzalez13,
  lin12}; see also the compilation in \citealt{planelles13}). This suggests that baryons are more broadly
distributed than the dark matter dominating the cluster potential 
  and a fraction of them may have been lost from (or never
  incorporated in) clusters, and especially so in groups and the least
  massive clusters (e.g., \citealt{renzini93}).
The same figure also shows in blue the fraction of the baryons which
are now in stars, i.e., $\m*/(\m* +M_{\rm gas})$, for both samples of
clusters.  A systematic decline of the star fraction with increasing
cluster mass is common to both samples. Notice that among clusters
with $M_{500}\simeq 10^{14}\,\msun$ this fraction is in the range
$\sim 0.2-0.3$, not too different from the value 0.17 adopted in Section
\ref{history} and substantially higher than the 0.1 value preferred by
\cite{loewe13}.  However,  the most massive clusters appear to  be
characterized by much smaller values, down to $\sim 0.06$.

\section{(Im)Possible Interpretations: The Conundrum}
\label{conundrum}
Clusters up to $\sim 10^{14}\,\msun$ apparently pose no serious
challenge. Their empirical iron yield is between $\sim 1$ and $\sim 2$
solar, within the expected range from supernova yields, and their iron
share is also between $\sim 1$ and $\sim 2$,
as known since a long time. Still, their baryon fraction is
appreciably below the cosmic value, indicating that the missing
baryons were never incorporated within the halo now hosting the
clusters, or were ejected (i.e., beyond $r_{500}$ for this work) from it under the action of some feedback. In
the latter case, metals may have been ejected as well, along with the
rest of the baryons, hence the above empirical iron yields should be
regarded as lower limits.

The problem arises from the more massive clusters, which apparently
demand yields well above solar and an iron share dramatically in
favor of the ICM. We emphasize again that the problem does not arise
uniquely from the two most massive clusters shown in Figure
\ref{goods} and Figure \ref{yields}, but instead it does from the
apparently well established trends with cluster mass of the gas fraction, the stellar
fraction and the metallicity, all illustrated in Figure
\ref{goods}. The {\it comundrum} arises from the factor of several drop of the stellar
fraction [$\propto \lr/(\m* + M_{\rm gas})$] with increasing cluster
mass which would demand a corresponding drop in metallicity, whereas the metallicity appears
to be constant among all clusters. 

We now list, in casual order, possible solutions of this conundrum.
\begin{itemize}
\item
In the most massive clusters there are several times more stars
out of galaxies than inside them, or, equivalently, the intracluster
light exceeds by such factor the luminosity of all the cluster
galaxies. Against this possible solution is lack of any 
evidence for such missing intracluster light (e.g.,
\citealt{zibetti05,andreon10,giallongo13}).

\item
The slope of the IMF above $\sim 1\,\msun$  tightly correlates with the
present mass of the clusters, i.e., {\it not} with the mass of the
galaxies as occasionally invoked (e.g., \citealt{cappellari12}).
Star-forming clouds at $z\gsim 2$
should know in advance the mass of the clusters in which their
products will be hosted $\gsim 10$ Gyrs later. As a kind of last
resort, IMF systematic
variations have been often invoked to fix problems, and this may be
one more example. However, this would demand that galaxies of given
mass would have experienced different IMFs in clusters with different
mass,
hence there should be systematic cluster-to-cluster differences in the
galaxy properties for which there is no evidence. For example, cluster
early-type galaxies follow closely the same fundamental plane
relation, irrespective of the cluster mass (e.g., \citealt{renzini06}
and references therein).
\item
As a variant to the above option one may think that the special (flat)
IMF is a specific property of the BCGs in the most massive clusters,
as in some of them massive starburst may be fed from intermitted 
cooling catastrophies of the ICM, hence representing a different star
formation mode \citep{mcdonald12}. However, especially among the most
massive clusters, BCGs account for only a small fraction of the total
cluster light and stellar mass. Hence, their IMF should be really
extreme for them to dominate the metal production of a whole cluster.
For example, in the two most massive clusters of our primary sample
the BCGs account for only up to $\sim 15\%$ of the total cluster
light, hence their metal yield should be $\gsim 20$ times solar for them to
account for most of the cluster metals.
\item
The yield is universal and extremely high ($\sim 4\,\zsun$)  but only the
most massive clusters have retained nearly all the metals, with most
metals having instead been lost by other, less massive
clusters. Formally, this might be accompleshed by an IMF substantially
flatter than Salpeter above $\sim 1\,\msun$ and/or with a subtantially
higher SNIa productivity $k_{\rm Ia}$ than reported in Section
\ref{history}. This may not contradict other evidences on cluster
galaxies, but it remains an {\it ad hoc} fix with no independent
evidences favoring it.
\item
The apparent lack of a correlation between ICM metallicity and the
stellar fraction may even suggest that the stellar populations of cluster galaxies have little to
do with the production of the metals now dispersed in the ICM (e.g.,
\citealt{loewe01,bregman10,loewe13,morsony14}). In this view, metals may have been
produced by an early stellar generation of very massive and/or Population
III stars, leaving
virtually no present-day low mass star counterparts. Like the others, also this  solution to
the conundrum lacks  independed observational evidence and would
actually require some cosmic conspiracy to ensure that ICM elemental
ratios turn out nearly solar, as observed (see \citealt{kyoko13} and
references therein), in spite of a radically different star formation
mode  compared to
the Galactic disk. Moreover, were Population III the solution
to the conundrum, this would imply an extremely high clustering
of Pop. III star, as such high metal pedestal is found only in
the most massive clusters. Extremely low metallicity stars are
indeed quite common within the Local Group, hence by comparison
with the clusters should have escaped virtually any pollution
from Pop. III stars. 

\item
As none of the above solutions is easy to accept, we must retain the
option that some of the observations illustrated in Figure \ref{goods}
may be faulty. The two most critical ones for the generation of the
cunundrum
are the drop in $\lr/M_{500}$ coupled to the constancy of metallicity. We have
already mentioned that \cite{russi} find a constant run of
$\lr/M_{500}$,
which automatically avoids the conundrum altogether. But we have
noticed that their measure of $M_{500}$ is of lower quality, being
obtained from the cluster mass-richness  relation instead than from modelling the X-ray
surface brightness distribution. The
alternative is that the measured metallicity is systemathically
overstimated with increasing cluster mass, which appears rather implausible
given the quality of the current X-ray data. 
\end{itemize}

In summary, none of these  possible solutions of the
conundrum appears attractive to us, hence we are left with no solution at all.

\begin{figure}
\includegraphics[width=84mm]{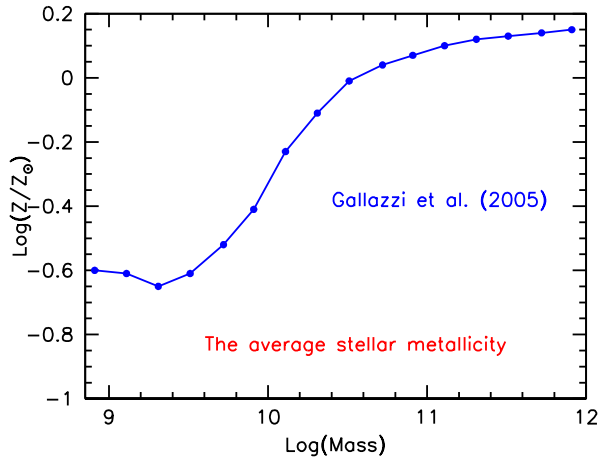}
\caption{The stellar metallicity vs. stellar mass relation for local
  galaxies as from Table 2 in Gallazzi et al. (2005)}
\label{gallazzi}
\end{figure}

\section{Past Metal Production as a Function of Galaxy Mass }
\label{winds}
In this section we further explore some consequences of the finding
that there is at least as much iron dispersed in the ICM as there 
is still locked into the cluster  stars and galaxies. In particular,
having empirically estimated the metal yield ($y$) we aim to
estimate the relative contribution of cluster galaxies to the metals
now in the ICM, as a function of galaxy mass, thus setting
constraints on the amount of metals individual galaxies should have lost.

It is well established that the metallicity of galaxies is an
increasing function of their stellar mass, both in the stellar as well as in
the gas components, both locally as well as at high redshift. From
the SDSS database, \cite{tremonti04} and \cite{gallazzi05} have
derived  the stellar mass-metallicity relation for the ISM and for
stars of local galaxies,  respectively. At higher redshifts the (ISM) mass-metallicity
relation shifts to lower metallicities (e.g. \citealt{erb06}),
possibly becoming steeper as a function of mass \citep{zahid13}.

Figure \ref{gallazzi} shows the stellar metallicity of local galaxies
as a function of stellar mass, as from Table 2 in \cite{gallazzi05}.
A similar flattening at high masses is also present when considering
the ISM metallicity, and \cite{tremonti04} interpreted it as the most
massive galaxies being able to retain virtually all the metals
produced by stars in the course of all previous evolution, whereas
lower mass galaxies would have lost in a wind (a major) part of them.  However, at
high redshifts evidence for galactic winds is ubiquitous for
star-forming galaxies of all masses (e.g.,
\citealt{pettini00,newman12}), with a mass loading factor (= ratio of the
mass loss rate to the star formation rate, SFR) of order of unity or
higher (\citealt{newman12,lilly13}). 
Thus, even the most massive galaxies must have lost a substantial
fraction of their metals, at least during their early evolution when both
their mass and SFR were growing rapidly \citep{renzini09,peng10}.
We actually interpret the asymptotic plateau at high masses as a
  result of galaxies turning passive due to {\it mass quenching}
  of star formation \citep{peng10}.  Thus, in this section we try
    to quantify how much metal mass should have been lost even by
the most    massive galaxies in order to achieve a metal-mass 
share above unity.

In this context one can also introduce the concept of {\it metal mass
  loading factor} of a galaxy (somehow analog to the mass loading
factor mentioned above), defined as the ratio of the metal mass
lost to the ICM/IGM to the metal mass still locked into its stars.
This quantity is then estimated below, for two values of the assumed
metal yield.

Here we first assume that the most massive galaxies have retained all the
metals that have been produced by the stars now in them, derive from
this the implied metal yield and check
what would be the resulting metal share between the ICM and
galaxies. We then relax this assumption to derive constraints on the
metal loss from galaxies if a metal share of order of unity (or
higher) is to be achieved, as demanded by the observations
(cf. Section \ref{history} and Section \ref{results}). We also assume
that the global metal yield $y$ is independednt of the metallicity of
the parent stellar population, as indeed indicated by theoretical
nucleosynthesis (e.g., \citealt{nomoto13}) according to which $y$ is a
weak function of metallicity. We then proceed to calculate what are
the fractions of the overall metal production that is now locked into
galaxies and that of the metals which have been ejected, both as a
function of galaxy mass.

For
the mass function of local galaxies we adopt the multi-Schechter
fits from \citet{peng10}, with
\begin{equation}
\phi(M)=\phi_{\rm B}(M)+\phi_{\rm R}(M),
\label{phi}
\end{equation}
being the sum the mass function of blue (star-forming)  and red
(quenched) galaxies, with
\begin{equation}
\phi_{\rm B} (M)= \phi^*_{\rm B}\left(M\over M^*\right)^{-1.4} e^{-M/M^*},
\label{phib}
\end{equation}
and
\begin{equation}
\phi_{\rm R} (M)= \phi^*_{\rm 1R}\left(M\over M^*\right)^{-0.4} e^{-M/M^*} +
\phi^*_{\rm 2R}\left(M\over M^*\right)^{-1.4} e^{-M/M^*},
\label{phir}
\end{equation}
where from Table 3 in \cite{peng10} we have $M^*=10^{10.67}\,\msun$,
$\phi^*_{\rm B}=1.014$, $\phi^*_{\rm 1R}=3.247$ and $\phi^*_{\rm
  2R}=0.214$.
Masses denoted with $M$ are here intended to be ``stellar masses'' of individual galaxies,
and we omit the subscript ``stars'' for simplicity.

\begin{figure}
\includegraphics[width=84mm]{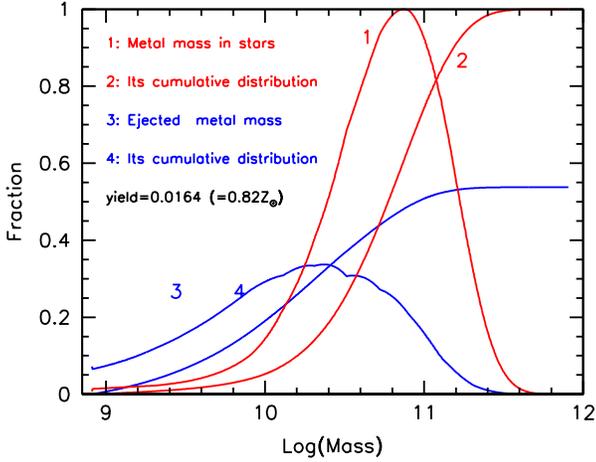}
\caption{The normalized distributions of the metals inside galaxies
  and ejected by them, curves 1 and 3, respectively, and the
  corresponding cumulative distributions, curves 2 and 4. These plots
  refer to an assumed metal yield $y=0.82\,\zsun$, corresponding to
  the assumption of most massive galaxies having evolved as a closed box}
\label{closed}
\end{figure}

\begin{figure}
\includegraphics[width=84mm]{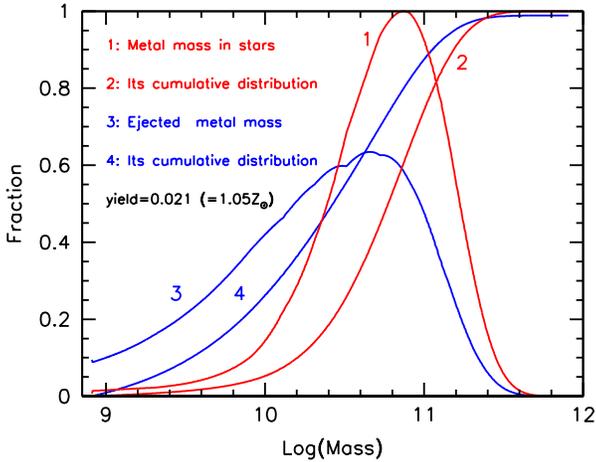}
\caption{The same as Figure \ref{closed} but for $y=1.05\zsun$, which
allows metal losses also from the most massive galaxies and ensures an
equal share of metals between the ICM and cluster galaxies}
\label{open}
\end{figure}

By its definition, the
total metal yield is given by the ratio of the total metal production
over the mass of gas that went into stars:
\begin{equation}
y={M_{\rm Z}^{\rm tot}\over M_{\rm tot}(0)}
\end{equation}
 where $M_{\rm tot}(0)$, the total mass turned into stars, is given by:
\begin{equation}
M_{\rm tot}(0)={1\over R}  \int_{\rm m} ^{M}M\,\phi(M)\,dM,
\label{mass}
\end{equation}
where $R$ is the average residual mass fraction, once taking
into account the mass return from dying stars. The value of $R$
depends on the actual star formation history of individual galaxies and its
accurate estimate is beyond the scope of this paper. By adopting
$R=0.58$ in Section \ref{implications} it was assumed that  the bulk
of stars in cluster galaxies are $\sim 11$ Gyr old, hence all galaxies
have the same $R$, and in this section
we stick on this assumption. Hence the relation $M(0)=1.72M$ holds
when referring to the total stellar mass of individual galaxies as well as
to the stellar mass of the galaxy population of a whole  cluster.

We now estimate the mass of metals inside galaxies and
outside them, as implied by the stellar mass-metallicity relation (MZR) of
\cite{gallazzi05} and an assumed value for the yield. This MZR does
not distinguish between clusters and field, hence we assume that it
applies to both environments. However, evidence exists for the
increase of the gas-phase metallicity with local overdensity for
star-forming satellite galaxies \citep{peng14}.
The mass of metals contained in galaxies up to mass $M$
is given by:
\begin{equation}
M_{\rm Z, stars} (M) = \int_{\rm m} ^{M}M\, Z(M)\,\phi(M)\,dM,
\label{mzstars}
\end{equation}
where $m$ is the minimum mass we are considering, say $\sim
10^9\,\msun$. Having assumed that the yield is independent of mass and
metallicity, the mass of metals that a galaxy of mass $M$ must have
ejected is given by the total production [$\equiv yM(0)$] minus the
metals still in the galaxy [$=Z(M)M$] and therefore the mass of metals produced by the same galaxies that are not locked into stars (i.e, that are either
in the ISM of individual galaxies or ejected into the intergalactic
medium, IGM) is given by:
\begin{equation}
M_{\rm Z, out} (M)= \int_{\rm m} ^{M}M\, [y/R-Z(M)]\,\phi(M)\,dM.
\label{mzout}
\end{equation}
The total mass of metals is therefore given by the sum of these two
integrals extended to the full mass range  ($\sim 10^{12}\,\msun$),
i.e., $M_{\rm Z}^{\rm tot}=M_{\rm Z,stars}+M_{\rm Z,out}$.
Among local galaxies the gas
fraction is a deacreasing function of mass, dropping from $\sim 30\%$
in $10^{10}\,\msun$ galaxies to $\lsim 10\%$ in $10^{11}\,\msun$
galaxies (e.g., \citealt{magdis12}). In the
following discussion we neglect the metals contained in the ISM of
individual galaxies, as they make a marginal contribution to the
global metal budget in the local Universe.

Finally, following its definition, the metal mass loading factor
$\lambda_{\rm Z}(M)$ is given by:
\begin{equation}
\lambda_{\rm Z}(M) = {y/R-Z(M)\over Z(M)}
\label{loading}
\end{equation}

We first assume that the most massive galaxies 
have completely retained all the metals that they have
produced. This allows us to estimate the yield as:
\begin{equation}
y={\zmax M\over M(0)} = 0.0164 = 0.82\zsun,
\label{yclosed}
\end{equation}
having taken $\zmax=0.0283$ from Figure \ref{gallazzi} and using
$\zsun=0.02$.
Correspondingly,
Figure \ref{closed} shows how the mass of metals is distributed among
galaxies as given by the integrand of Equation \eqref{mzstars} whereas
the cumulative distribution $M_{\rm Z,stars} (M)$ is given by the
same equation. Moreover, the figure also shows how galaxies in the
various mass bins have contributed to the mass of metal now out of
stars,
as given by the integrand of Equation \eqref{mzout}, and the
corresponding cumulative distribution. For display purposes, the plots
relative to the metals within stars have been normalized to unity, but
those relative to metals not locked into stars maintain the proper
proportion with respect to the former ones.

Several interesting aspects are self-evident from Figure
\ref{closed}: most of the stellar metals are contained in galaxies with
$M\sim M^*$ whereas the bulk of ejected metals comes from somewhat
lower mass galaxies, as revealed by the peak of curve 3 being shifted to lower masses with thespect to the peak
of 
curve 1 (see also \citealt{thomas98}). Still, most of the action is due to galaxies
which today are more massive than $\sim 10^{10}\,\msun$. Perhaps
more importantly, under the assumption that the most massive galaxies
do not eject any metals, the mass of metals ejected is about half of
the mass of metals still locked into galaxies: i.e., $\sim 2/3$ of the
metals are in stars and $\sim 1/3$ are dispersed  outside galaxies in
the IGM (with a minor fraction still in the ISM).

This $\sim 0.5$ share of metals between the IGM/ICM and stars falls
somewhat short of the $\gsim 1$  share that is found in
clusters of galaxies, as reported in Section \ref{history} and
illustrated  in Figure \ref{yields} and \ref{yieldt}.  
In order to achieve
a higher share, more in favor of the IGM/ICM, one has to assume a
higher metal yield, i.e., a higher value $y$ in Equation \eqref{mzout}.
After a few tentatives we find that a value $y=0.021 \quad
(=1.05\zsun)$ gives a nearly fifty-fifty share of metals between IGM/ICM and stars and the
corresponding distributions are shown in Figure \ref{open}. Having
allowed also massive galaxies to lose metals, the mass of the peak
metal producers is now higher than in the former case, quite close
indeed to $M*$. 

It is somehow surprizing (and encouraging) that just with $y=\zsun$ one gets a metal
share of unity, not far from the values exhibited by the clusters, at
least those of mass up to $\sim 10^{14}\,\msun$. Compared to the
previous case, the contibution to metals outside galaxies has nearly
doubled at all masses, and more so towards the high mass end.
One can conclude that the observed $\gsim 1$ metal share in clusters
requires $y\gsim\zsun$, hence that also the most massive
galaxies had to eject a substantial fraction of the metals they have
produced.
Still, the most massive galaxies do not contribute much to
the ICM metals because they are very rare, even if the mass of metals
each of them has ejected, i.e., $[1.72y-Z(M)]M$,  is actually maximal.

\begin{figure}
\includegraphics[width=74mm]{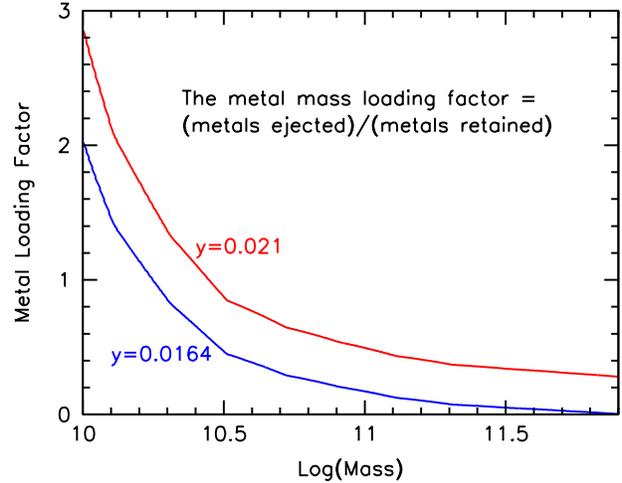}
\caption{The metal mass loading factor as a function of stellar mass
  for the two indicated values of the metal yield}
\label{loadingf}
\end{figure}

Finally, Figure \ref{loadingf} shows the mass dependence of the metal
mass loading factor from Equation \eqref{loading} for the two values of the yields used above. By
construction, in the case of the lower value of $y$ the loading factor
vanishes towards high masses. For the higher value of $y$ the loading factor
is still $\sim 0.3$ even at the highest masses, i.e., $\sim 1/4$ of
the produced metals are ejected and $\sim 3/4$ are retained by
such galaxies. Of course, for the higher iron share of the massive
clusters a higher yield is required, implying that even the most massive galaxies would have lost the majority of the metals
they have produced.

\section{Conclusions}
\label{end}
We have revisited the metal budget of clusters of galaxies using
recent cluster data for a sample of clusters for which all four
  basic parameters are homogeneously measured
within consistent radii, namely core-excised, mass-weighted
metallicity plus total, stellar and ICM masses. We further use a wider sample
  of clusters for which one (or two) such parameters are not
  available, but for which the available data are of high
  quality. Together, these various samples concur in establishing the
  trends among the four cluster parameters that we discuss in this paper.

For clusters with mass $M_{500}$ up to $\sim 10^{14}\,\msun$
the total mass of metals is well within the limits expected from
standard semi-empirical nucleosynthesis, and a metal yield $y\simeq
(1-2)\zsun$.
However, when considering more massive clusters a sizable drop of the
cluster stellar luminosity per unit cluster mass ($\lr/M_{500}$) is
not accompanied by a drop of the ICM metallicity, as expected
if the metal yield is constant. Conversely, the empirical metal yield
appears to increase to many times solar for the most massive clusters
approaching $M_{500}=10^{15}\,\msun$. 

Various possible solutions to this {\it conundrum} are discussed,
either appealing to missing intracluster light, or systematic
cluster-to-cluster differences in the IMF, or of a universal, yet very
high metal yield, or even invoking an extinct population of massive
stars unrelated to the stellar populations still shining today.  
Some of such
hypothetical solutions appear to be rather astrophysically implausible.
Others, although plausible, are not supported by independent
evidences, hence remain {\it ad hoc}. For these reasons
we refrain from favoring any of them. We still
cannot exclude  the possibility of some systematic bias affecting even
the
current best measurements of some of the basic cluster parameters, such as
their total and ICM mass, the cluster stellar luminosity and the ICM
metallicity, but we are not able to identify any obvious bias in the
data. We just emphasize that the most serious problem arises from the
most massive clusters, with $M_{500}\gsim 5\times 10^{14}\,\msun$,
which then would deserve focused observational efforts. Our wish
is that this paper could help triggering such efforts.

We include in our study an attempt at estimating the mass of metals
that galaxies of given mass must have ejected in the course of their
evolution, based on the local stellar mass-metallicity relation and an
assumed universal yield (i.e., assuming that the galaxies
we see today have produced all the metals). Under these assumptions, it is found that a
yield $y=\zsun$ ensures a nearly 50-50 share of metals between the
IGM/ICM and stars, close to what observed for typical
$M_{500}=10^{14}\,\msun$ clusters.  However, for the much
higher iron share of the more massive clusters a yield at least
twice solar would be required, implying that even the most massive
glaxies whould have lost the majority of the metals that
have been produced.

\section*{Acknowledgments}
We are grateful to M. Sun for having provided us with his measurement
of the iron abundance in the cluster A2092 and to A.  Gonzalez for a
useful discussion on the likely origin of
the offset of the $\lr/M_{500}$ ratios apparent in Figure \ref{goods}. 
AR acknowledges support from the INAF-PRIN 2010.

\label{lastpage}

\end{document}